# Fiber-comb-stabilized light source at 556 nm for magneto-optical trapping of ytterbium

Masami Yasuda,[1,2] Takuya Kohno,[1,2] Hajime Inaba,[1,2] Yoshiaki Nakajima,[1,2] Kazumoto Hosaka,[1,2] Atsushi Onae,[1] and Feng-Lei Hong,[1,2]

[1]*National Metrology Institute of Japan (NMIJ), National Institute of Advanced Industrial Science and Technology (AIST), Central 3, 1-1-1 Umezono, Tsukuba, Ibaraki 305-8563, Japan*

[2]*CREST, Japan Science and Technology Agency, 4-1-8 Honcho Kawaguchi, Saitama 332-0012, Japan*

A frequency-stabilized light source emitting at 556 nm is realized by frequency-doubling a 1112-nm laser, which is phase-locked to a fiber-based optical frequency comb. The 1112-nm laser is either an ytterbium (Yb)-doped distributed feedback fiber laser or a master-slave laser system that uses an external cavity diode laser as a master laser. We have achieved the continuous frequency stabilization of the light source over a five-day period. With the light source, we have completed the second-stage magneto-optical trapping (MOT) of Yb atoms using the $^1S_0$ - $^3P_1$ intercombination transition. The temperature of the ultracold atoms in the MOT was 40 μK when measured using the time-of-flight method, and this is sufficient for loading the atoms into an optical lattice. The fiber-based frequency comb is shown to be a useful tool for controlling the laser frequency in cold-atom experiments.

## 1. Introduction

At the end of the last century, mode-locked femtosecond lasers were demonstrated for measuring laser frequencies [1,2]. An optical frequency comb obtained by injecting the light of a mode-locked Ti:sapphire laser into a photonic crystal fiber [3] can cover more than one octave in frequency, enabling the direct measurement of the carrier-envelope offset frequency of the comb [4] and hence the absolute frequency of a laser. Optical frequency combs have revolutionized research on optical frequency measurement and are now utilized for various applications including optical clocks [5-7], ultrafast phenomena [8], and astronomical observations [9].

An optical frequency comb is not only a useful tool for optical frequency measurement but also an excellent way of controlling laser frequency. Traditionally, laser frequency control was realized by using frequency references, such as a stable optical cavity and atomic or molecular transitions. An optical cavity is relatively easy to make as a frequency reference but it has the demerit of frequency drift. Also, since the coating bandwidth is usually limited to a specific wavelength, different optical cavities are needed for different wavelengths. As for atomic or molecular transitions, the spectral linewidth may not meet the requirement as a frequency reference. It may also be very difficult to detect the signal from a beam system when using an isotope with little natural abundance. Again, each transition provides one reference at a specific wavelength. However, with the optical frequency comb as a frequency reference, the wavelength range can exceed one octave and it derives its stability directly from a frequency standard. The drawbacks of the frequency comb were its poor reliability over long-term operation and its cost. Recently developed frequency combs based on mode-locked erbium-doped fiber lasers (fiber combs) achieve turnkey use [10] and robust long-term operation [11].

Furthermore, the fiber comb uses parts commonly employed for telecom applications that are relatively inexpensive.

Frequency-stabilized lasers are essential for high-resolution spectroscopy, cold atom and quantum information experiments. Cold atom experiments with ytterbium (Yb) are of great interest for studies related to quantum many-body systems [12,13] and optical clocks [14-16]. The laser cooling of Yb atoms usually requires either two-stage magneto-optical trapping (MOT) using both the broad $^1S_0 - \,^1P_1$ and the narrow $^1S_0 - \,^3P_1$ transitions (see Fig. 1) or single-stage MOT directly using the narrow $^1S_0 - \,^3P_1$ (intercombination) transition [17]. Since the linewidth of the intercombination transition is 182 kHz, the light source for MOT has been prepared using frequency stabilization based on an ultra-low-expansion (ULE) cavity [18] or an Yb atomic beam [19,20].

In this paper, we demonstrate a frequency-stabilized light source at 556 nm based on a home-made fiber comb [11,21] for the MOT that uses the Yb intercombination transition. An Yb-doped distributed feedback (DFB) fiber laser is phase-locked to the fiber comb to realize both laser frequency jitter reduction and long-term frequency stabilization. This long-term operation is especially important for clock experiments because atomic clocks are compared with International Atomic Time (TAI) using a five-day segment in the comparison scheme. With the developed light source, we have achieved second-stage MOT using the intercombination transition, which was sufficient for loading the atoms into an optical lattice. We have also demonstrated that the present stabilization scheme is applicable to an external-cavity diode laser (ECDL) at 1112 nm. In this case, the ECDL linewidth is reduced with a tight phase lock between the ECDL and the fiber comb.

## 2. Experimental setup

Figure 2 is a schematic diagram of our experimental setup. The fiber comb contains an erbium-doped mode-locked fiber laser, amplifiers, highly nonlinear fibers, and detection parts. The fiber laser is a ring resonator that employs nonlinear polarization rotation as the mode-locking mechanism [11]. The spectral bandwidth and the repetition rate ($f_{rep}$) of the fiber laser were 60 nm and 122 MHz, respectively. The average output power of the laser was about 18 mW and was distributed equally to four parallel branches. Each branch contains an amplifier, a highly nonlinear fiber and a detection part for various applications. The amplifier generates an average output power of about 58 mW. The output continuum from the highly nonlinear fiber covers a wavelength range exceeding one octave (1000 to 2050 nm). The first branch was used to detect the carrier-envelope offset frequency ($f_{CEO}$) [4] and $f_{rep}$. The second branch was used to detect the beat frequency ($f_b$) between the DFB fiber laser and the comb modes at 1112 nm. The third and fourth branches are currently unused and will be employed in the future for other lasers required for the experiment. The multi-branch configuration enables us to obtain $f_{CEO}$ and $f_b$ with a high and stable signal-to-noise ratio (S/N). The detected $f_{rep}$ and $f_{CEO}$ were phase locked to an H-maser by feedback controlling the length and pump power of the fiber laser oscillator, respectively. In this way, the entire fiber comb was phase locked to the H-maser, and the absolute frequency of the $n$th comb mode ($f_n$) could be calculated as $f_n = n\, f_{rep} + f_{CEO}$. Our fiber comb system is described in detail elsewhere [11,21,22].

The 1112-nm DFB fiber laser (commercially available from Koheras) has a linewidth of ~ 50 kHz. The laser emits about 200 mW of output power in a polarization-maintaining (PM) optical fiber. A PM fiber coupler was used to divide the fiber laser output into two parts. 1 % of the laser power was used for a heterodyne beat measurement with the fiber comb. The detected $f_b$ was also phase locked to the H-maser by feedback controlling the length of the DFB fiber laser through a piezoelectric transducer actuator (PZT) installed in the fiber laser. In this way, the DFB fiber laser was phase locked to the fiber comb. $f_b$ was monitored with a spectrum analyzer and measured by using a dead-time-free frequency counter. The time base of the frequency counter was connected to the H-maser.

The remaining 99 % of the DFB fiber laser power was used for second harmonic generation (SHG) with a periodically-poled-lithium-niobate (PPLN) waveguide (WG) device. The PPLN-WG device was designed to have a PM fiber pigtail so that the output from the fiber coupler was directly connected to the fiber pigtail, thus realizing stable long-term operation. The generated 556-nm light was collimated by using an objective lens and was separated out from the original 1112-nm light by using a dichroic mirror. A half-wave plate was used to rotate the beam polarization plane such that a polarization beam splitter (PBS1) divided the 556-nm beam with an appropriate ratio into vertical and horizontal beams for the MOT [20].

As a backup system for the DFB fiber laser, we have also built a 1112-nm ECDL as a master laser. An optical power of about 150 mW was achieved by injection-locking a slave laser with the master laser. A Faraday rotator was used to arrange the laser polarization so that the master and slave laser lights could be separated by PBS2 in Fig. 2. The combination of the Faraday rotator, PBS2 and PBS3 formed an isolator to prevent the slave laser from any possible back-reflected light. Part of the master laser light was used for the heterodyne beat measurement with the fiber comb. The slave laser light was coupled into a PM fiber and connected to the PPLN-WG device.

## 3. Experimental results

The fiber comb was phase locked to the H-maser to convert the frequency characteristics of the H-maser to optical frequencies. The frequency instability of our H-maser is about $4\times10^{-13}$ at an averaging time ($\tau$) of 1 s. This corresponds to a frequency instability of about 100 Hz for the 1112-nm light. For $\tau \sim 1$ ms, the frequency instability of the H-maser is very large and corresponds to a frequency instability of above 100 kHz for the 1112-nm light. On the other hand, the measured free-running linewidth (for $\tau \sim 1$ ms) of the fiber comb modes was < 10 kHz [23]. If we tightly lock the comb to the H-maser, the larger frequency noise of the H-maser will be transferred to the fiber comb in the ms time range. The strategy was to adjust the time constant of the feedback loop to maintain the comb linewidth but reduce the frequency jitter and drift using the H-maser.

Figure 3 shows the beat frequencies between the DFB fiber laser and the fiber comb modes observed with the spectrum analyzer, when the DFB fiber laser was phase locked to the fiber comb. The resolution bandwidth of the spectrum analyzer was 30 kHz. $f_b$ is the beat frequency between the fiber DFB laser and the nearest comb mode. ($f_{rep}$ - $f_b$) is the beat frequency between the DFB fiber laser and the second nearest comb mode. The S/N of the observed beat signals was about 35 dB at a resolution bandwidth of 30 kHz, and was sufficient for the phase-lock loop. Owing to the limited servo bandwidth obtained with the PZT in the DFB fiber laser, the phase locking of the DFB fiber laser to the fiber comb was not tight enough to reduce the linewidth of the DFB fiber laser (~ 50 kHz) to that of the fiber comb (< 10 kHz). Therefore, the observed linewidth of $f_b$ was basically the free-running linewidth of the DFB fiber laser. However, the frequency jitter and drift were reduced to that of the H-maser through the fiber comb. The linewidth of the fiber-comb-stabilized DFB fiber laser at 1112 nm is doubled to ~ 100 kHz at 556 nm and were good enough for the MOT using the intercombination transition (linewidth = 182 kHz). Figure 4 shows the variation in $f_b$ measured with a dead-time-free frequency counter, when $f_b$ was phase-locked to the H-maser. The counter gate time was 1 s. The inset in Fig. 4 shows the Allan deviation [24] calculated from the measured beat frequency. The $1/\tau$ character of the observed Allan deviation indicates successful phase locking between the DFB fiber laser and the fiber comb.

As an alternative to the DFB fiber laser, we have also introduced a master-slave laser system using an external cavity diode laser as a master laser. The free-running linewidth of the ECDL was estimated to be about 300 kHz by observing the beat signal between the ECDL and the fiber comb modes. This linewidth was reduced to < 10 kHz by realizing tight phase locking between the ECDL and the fiber comb using the LD current as a servo port. The achieved servo bandwidth was about 500 kHz. The tight phase locking was also confirmed by observing the in-loop beat signal between the ECDL and the fiber comb mode (as shown in Fig. 5). The observed linewidth of the in-loop beat signal was limited by the resolution bandwidth of the spectrum analyzer at 10 Hz.

The fiber-comb-stabilized laser sources were used for the second-stage MOT of Yb. With the PPLN-WG device, 5 mW of green light was generated from 50 mW of input fundamental light, which was not the maximum power of the DFB fiber laser or of the slave laser injected with the ECDL. It is worth noting that the perfect long-term operation of the single-pass PPLN-WG device has also contributed to the robustness of the system by comparison with a power build up cavity. One million atoms were transferred into the second-stage MOT from the first-stage MOT, where about ten million atoms were trapped [25,26]. The temperature of the ultracold atoms in the second-stage MOT was measured using the time-of-flight (TOF) method. Figure 6(a) shows the experimental setup for the TOF method. A 399-nm probe beam was set about 14 mm below the MOT and was 1 mm thick and 5 mm wide. After releasing the atoms from the MOT by turning off the trapping laser and the magnetic field, the atoms dropped and passed through the probe laser light. The 399-nm fluorescence from the atoms passing through the probe laser was collected by a lens and observed with a photodiode. Figure 6(b) shows the observed TOF signal of $^{171}$Yb (nuclear spin $I = 1/2$) atoms as a dashed curve. The atomic temperature is roughly proportional to the square of the TOF signal width. The TOF signal was fitted with a theoretical curve [27] (solid curve) giving a temperature of 40 μK for the $^{171}$Yb atoms in the second-stage MOT. The achieved temperature was sufficiently low for loading the atoms into an optical lattice formed by the standing wave of a cw Ti:sapphire laser (power ~ 400 mW, radius ~ 25 μm). We have performed a spectroscopic observation of $^{171}$Yb atoms in an optical lattice using a clock laser at 578 nm [28,29] and also measured the absolute frequency of the clock transition [15]. The second-stage MOT using other isotopes of Yb ($^{174}$Yb and $^{173}$Yb) has also been realized. The measured temperatures of the $^{174}$Yb ($I = 0$) and $^{173}$Yb ($I = 5/2$) atoms in the second-stage MOT were 130 and 15 μK, respectively. The temperature achieved in the MOT decreased as the nuclear spin increased [19].

## 4. Discussion

In the present experiment, the fiber comb was phase locked to an H-maser, which is a high-grade frequency standard and may not be available in all institutes. We have also performed the experiment using a commercial rubidium (Rb) clock as a time base instead of the H-maser. The second-stage MOT of Yb has also been realized using the Rb time base. The achieved temperature of the atoms did not depend on the choice of time base. The frequency instability of the Rb clock is about $3\times10^{-11}$ at $\tau =1$ s, which corresponds to a frequency instability of about 16 kHz for the 556-nm light and is much smaller than the linewidth of the intercombination transition.

We have demonstrated that a fiber-comb-stabilized light source can be successfully used for the MOT of Yb atoms. Whether or not this technique is applicable to other atoms depends on the relationship between the linewidths of the intercombination transition and the fiber comb mode. For example, the linewidth of the $^1S_0$ - $^3P_1$ intercombination transition of $^{87}$Sr is 7.6 kHz and is smaller than that of the fiber comb used in the present experiment. This means that we need to narrow the linewidth of the fiber comb for the Sr experiment. Recently, we have demonstrated that a fiber comb with an intra-cavity electro-optic modulator could transfer the linewidth from one laser to another precisely at the millihertz level [21]. With such a fiber comb, we should be able to transfer the linewidth of the clock laser, which is usually locked to an ultra-stable ULE cavity, to the laser source for the intercombination line MOT. In our new $^{87}$Sr lattice clock project, the light source for the intercombination transition will be phase locked to the clock laser using the narrow-linewidth fiber comb.

In the lattice clock experiment, many laser sources at different wavelengths are used for different purposes. As described in section 2, our fiber comb has four parallel branches, two of which are spare for other applications. We plan to use one of the spare branches to stabilize the frequency of the lattice laser at 759 nm (magic wavelength, no atomic or molecular resonance can be used). To achieve a better frequency uncertainty for the Yb lattice clock than that of the Cs clock, for example at $1\times10^{-17}$, the frequency of the lattice laser needs to be stabilized within an uncertainty of 600 kHz [15]. This requirement can be met by also phase locking the lattice laser to the fiber comb. Although the fiber comb does not directly cover the lattice clock wavelength, frequency doubling of the comb components around 1519 nm enables us to perform a beat frequency measurement between the lattice laser and the fiber comb at 759 nm. The multi-branch fiber comb is a useful tool for controlling the frequencies of different lasers in the cold-atom experiments.

For cold-atom experiments, useful wavelengths are found in the visible to near infrared range. The fiber comb directly covers an octave wavelength range from 1000 to 2050 nm. By using the SHG of the comb components, lasers with wavelengths ranging from 500 to 1000 nm can be easily measured or controlled. For wavelengths shorter than 500 nm, one possible approach is to use the broadened spectrum obtained by injecting the SHG of a mode-locked fiber laser into a photonic crystal fiber [30]. UV combs have been generated using high-harmonic generation at wavelengths shorter than 100 nm [31,32]. On the other hand, for wavelengths longer than 2000 nm, the sum-frequency generation [33,34] and the difference-frequency generation [35,36] of an optical frequency comb were demonstrated to be effective methods.

The robust long-term frequency stabilization of a laser source is important not only for cold-atom experiments but also for various applications including optical fiber transfer [37], precise spectroscopy, quantum information science, and gravitational wave detection. A fiber comb is a new and powerful tool for such applications. Research on fiber combs pursuing high precision, high reliability, low cost production, and commercialization is very active. We expect that in the near future, most optical laboratories will be equipped with an optical frequency synthesizer made using fiber combs.

In summary, we have developed a fiber-comb-stabilized light source for the second-stage MOT of Yb using the $^1S_0$ - $^3P_1$ intercombination transition. The light source operates continuously over five days, achieving both laser frequency jitter reduction and long-term frequency stabilization. Using the developed light source, we have completed the second-stage MOT of Yb, which was successfully used in an Yb optical lattice clock.

## Acknowledgments

We thank H. Katori for many enlightening discussions on optical lattice clocks. We are grateful to T. Fukuhara and M. Takamoto for helpful discussions on experiments. Part of this work was supported by the "Grant for Industrial Technology Research (financial support to young researchers)" program of the New Energy and Industrial Technology Development Organization.

© 2010 Optical Society of America, Inc.32. R. J. Jones, K. D. Moll, M. J. Thorpe, and Jun Ye, "Phase-coherent frequency combs in the vacuum ultraviolet via high-harmonic generation inside a femtosecond enhancement cavity," Phys. Rev. Lett. **94**, 193201 (2005).

33. O. D. Mücke, O. Kuzucu, F. N. C. Wong, E. P. Ippen, F. X. Kaertner, S. M. Foreman, D. J. Jones, L.-S. Ma, J. L. Hall, J. Ye, "Experimental implementation of optical clockwork without carrier-envelope phase control," Opt. Lett. **29**, 2806-2808 (2004).

34. J. Jiang, A. Onae, H. Matsumoto, and F.-L. Hong, "Frequency measurement of acetylene-stabilized lasers using a femtosecond optical comb without carrier-envelope offset frequency control," Opt. Express **13**, 1958-1965 (2005).

35. P. Malara, P. Maddaloni, G. Gagliardi, and P. De Natale, "Absolute frequency measurement of molecular transitions by a direct link to a comb generated around 3-μm," Opt. Express **16**, 8242-8249 (2008).

36. K. Takahata, T. Kobayashi, H. Sasada, Y. Nakajima, H. Inaba, and F.-L. Hong, "Absolute frequency measurement of sub-Doppler molecular lines using a 3.4-μm difference-frequency-generation spectrometer and a fiber-based frequency comb," Phys. Rev. A **80**, 032518 (2009).

37. F.-L. Hong, M. Musha, M. Takamoto, H. Inaba, S. Yanagimachi, A. Takamizawa, K. Watabe, T. Ikegami, M. Imae, Y. Fujii, M. Amemiya, K. Nakagawa, K. Ueda, and H. Katori, "Measuring the frequency of a Sr optical lattice clock using a 120 km coherent optical transfer," Opt. Lett. **34**, 692-694 (2009).
16

Figure captions

Fig. 1 (Color online) Energy levels of $^{171}$Yb. Wavelengths and natural linewidths are indicated for the relevant cooling, trapping, clock and probing transitions.

Fig. 2 (Color online) Schematic diagram of the fiber-comb-stabilized light source for the second-stage magneto-optical trapping (MOT) of Yb. DFB, distributed feedback; PM fiber, polarization-maintaining fiber; PPLN, periodically poled lithium niobate; WG, waveguide; PBS, polarization beam splitter; ECDL, external cavity diode laser; FR, Faraday rotator; $f_{CEO}$, carrier-envelope offset frequency; $f_{rep}$, repetition rate; $f_b$, beat frequency between the DFB fiber laser and the nearest comb mode with a mode number of n; $f_n$, frequency of n-th comb mode.

Fig. 3 Beat frequencies between the DFB fiber laser and the fiber comb modes at 1112 nm observed with a spectrum analyzer. The resolution bandwidth was 30 kHz. $f_b$ is the beat frequency between the DFB fiber laser and the nearest comb mode. $f_{rep}$ is the repetition rate of the comb. ($f_{rep}$ - $f_b$) is the beat frequency between the DFB fiber laser and the second nearest comb mode. The inset shows the expanded spectrum of $f_b$.

Fig. 4 Variation in the measured beat frequencies of the DFB fiber laser and the fiber comb mode obtained using a frequency counter, when the beat was phase-locked to a H-maser. The counter gate time was 1 s. Indicated data are the frequency deviation from average. A continuous 5-day phase lock was achieved. The inset shows the Allan deviation calculated from the measured beat frequency.

Fig. 5 Beat frequency between the ECDL and the fiber comb mode at 1112 nm observed with a spectrum analyzer, when the ECDL was tightly phase-locked to the fiber comb. The resolution bandwidth was 10 Hz.

Fig. 6 (Color online) Temperature measurement of ultracold atoms in magneto-optical trapping (MOT) using the time-of-flight (TOF) method. (a) Experimental setup for the TOF method. (b) TOF signal (dashed curve) of Yb atoms released from the second-stage MOT. The solid curve is a fitting of the signal giving a temperature of 40 μK for the Yb atoms in the MOT.

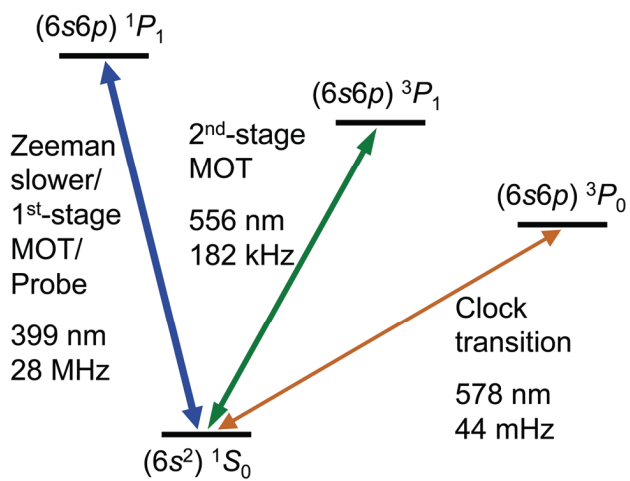

Fig. 1

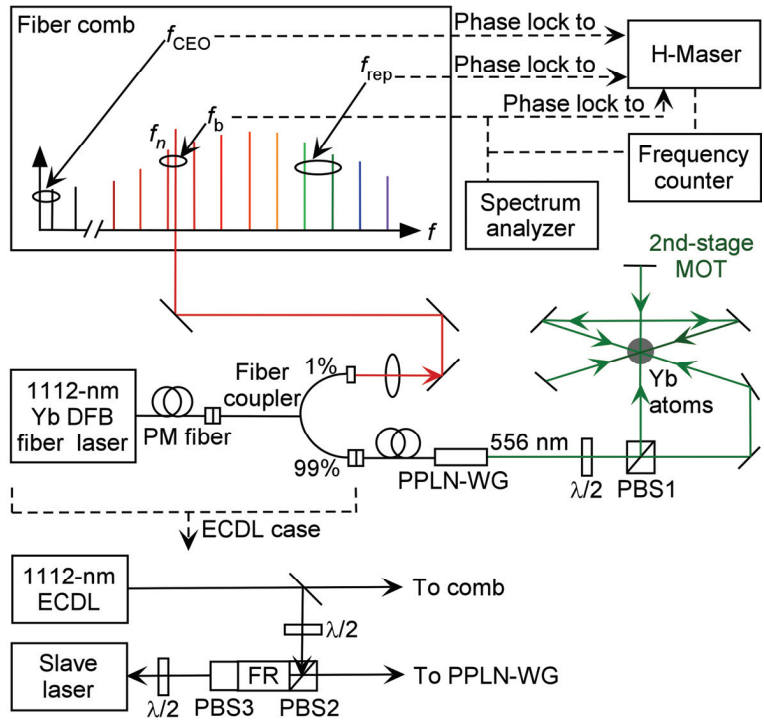

Fig.2

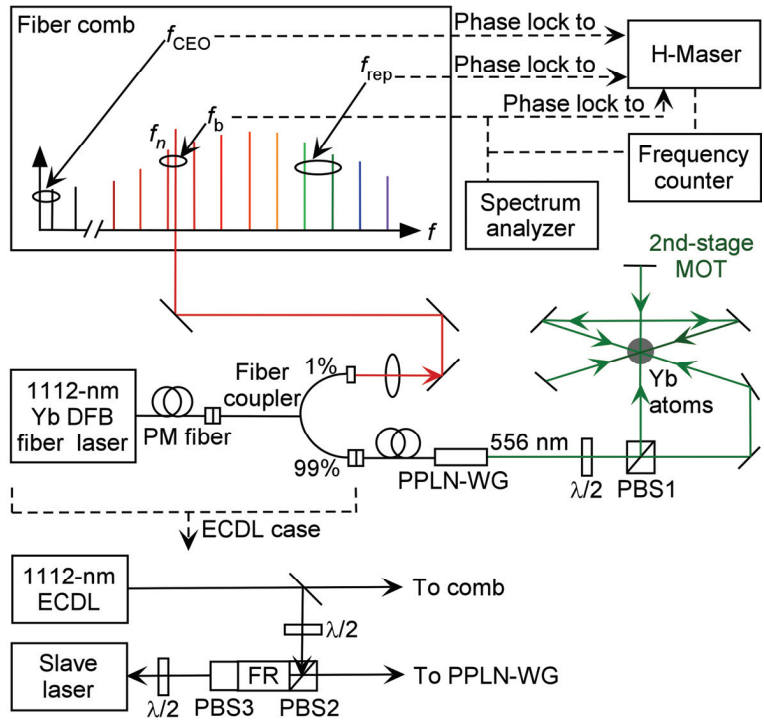

Fig.2

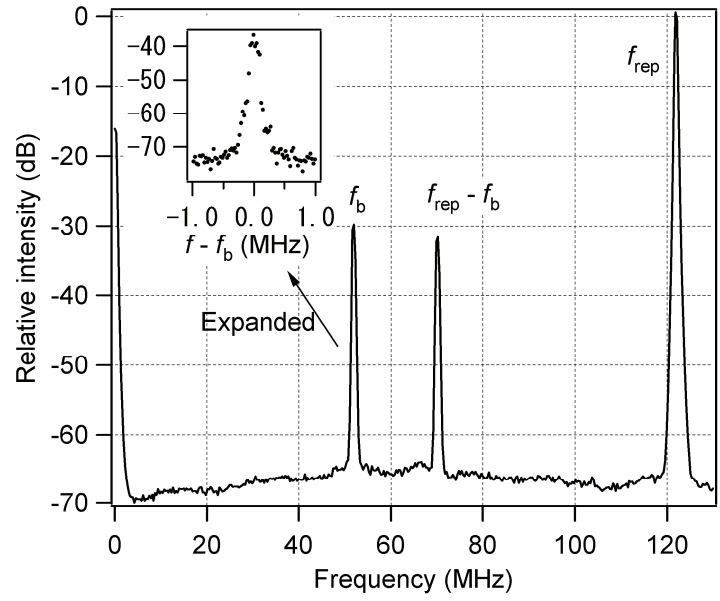

Fig. 3

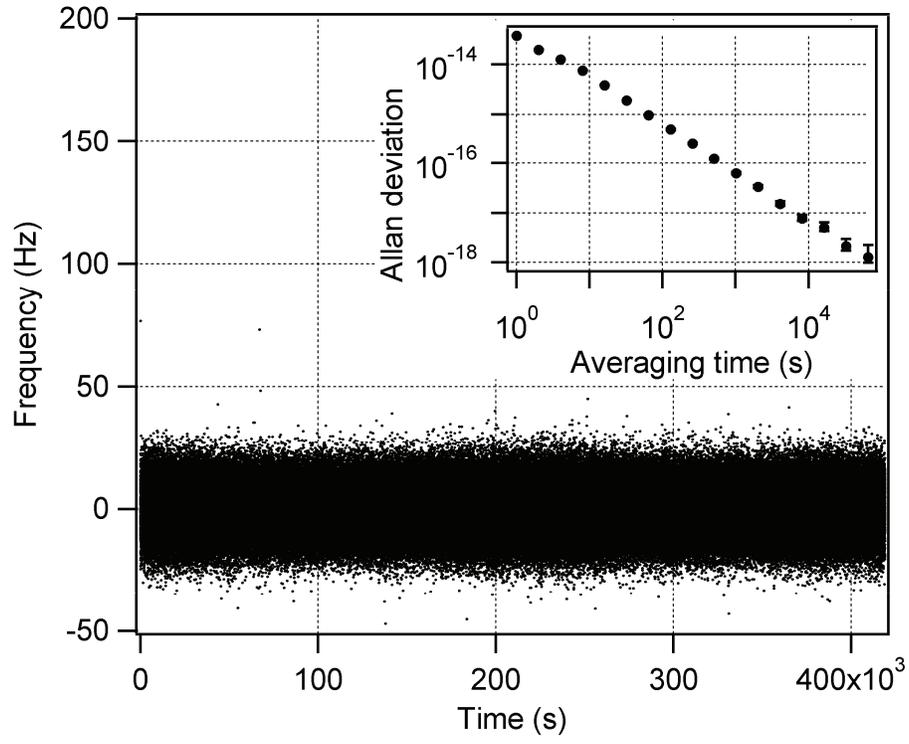

Fig. 4

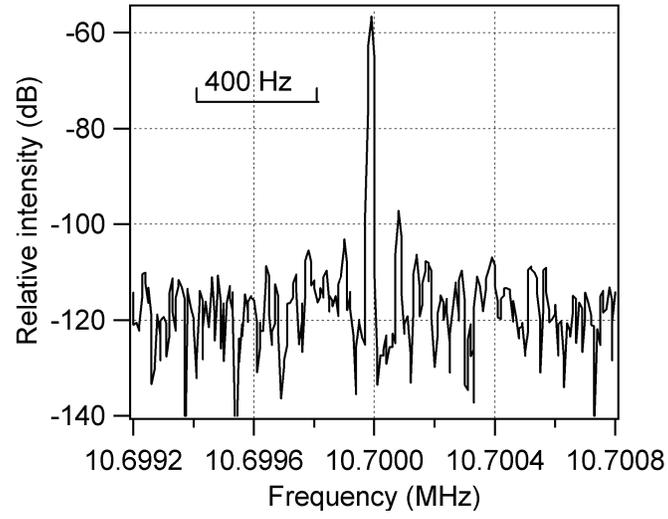

Fig. 5

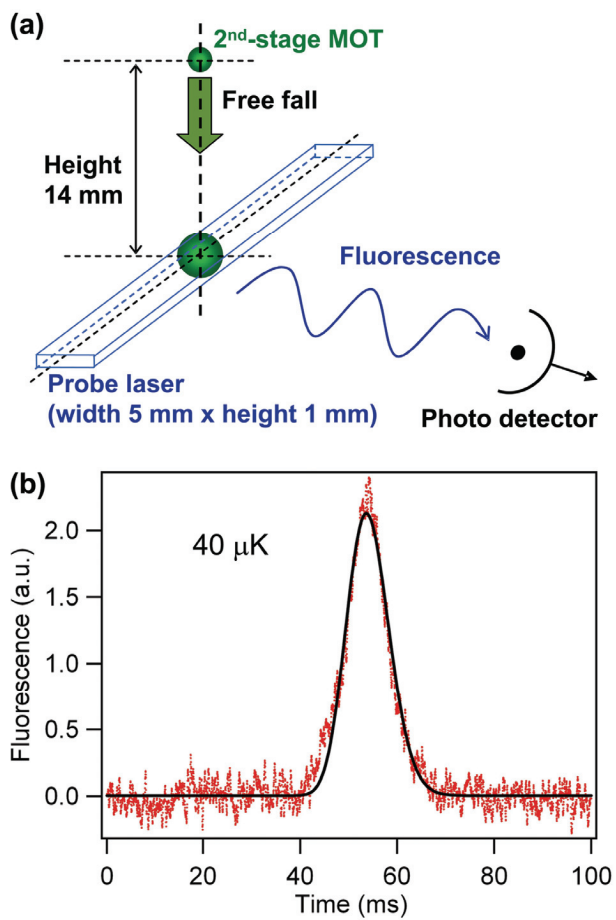

Fig. 6